\documentclass[12pt,preprint]{aastex}

\slugcomment{Submitted to the {\it Astrophysical Journal}}
\shorttitle{QS Vir: cataclysmic or hot air?}
\shortauthors{Matranga et al.}

\begin{document}

\title{Detection of accretion X-rays from QS Vir: cataclysmic or a lot of hot air?}

\author{Marco Matranga,\altaffilmark{1} Jeremy
  J.~Drake,\altaffilmark{1} Vinay Kashyap,\altaffilmark{1} Danny Steeghs,\altaffilmark{2} }
\affil{$^1$Harvard-Smithsonian Center for Astrophysics,
    60 Garden Street, Cambridge, MA 02138}
\email{mmatranga@cfa.harvard.edu}
\affil{$^2$Department of Physics, University of Warwick, Coventry CV4
  7AL, UK}

\begin{abstract} 
An XMM-Newton observation of the nearby ``pre-cataclysmic'' short-period ($P_{orb} = 3.62$~hr) binary QS Vir (EC~13471-1258)  revealed regular narrow X-ray eclipses when the white dwarf passed behind its M2--4 dwarf companion.  The X-ray emission provides a clear signature of mass transfer and accretion onto the white dwarf.  The low-resolution XMM-Newton EPIC spectra are consistent with a cooling flow model and indicate an accretion rate of  $\dot{M} = 1.7 \times 10^{-13} M_\odot$/yr.  
At 48~pc distant, QS~Vir is then the second nearest accreting cataclysmic variable known, with one of the lowest accretion rates found to date for a non-magnetic system. 
To feed this accretion through a wind would require a wind mass loss rate of $\dot{M}\sim 2\times 10^{-12}M_\odot$~yr$^{-1}$ if the accretion efficiency is of the order of 10\%.  Consideration of likely mass loss rates for M dwarfs suggests this is improbably  high and pure wind accretion unlikely.  A lack of accretion disk signatures also presents some difficulties for direct Roche lobe overflow.  
We speculate that QS~Vir is on the verge of Roche lobe overflow, and that the observed  mass transfer could be supplemented by upward chromospheric flows on the M dwarf, analogous to spicules and mottles on the Sun,  that escape the Roche surface to be subsequently swept up into the white dwarf Roche lobe.  If so, QS~Vir would be in a rare evolutionary phase lasting only a million years.
The X-ray luminosity of the M~dwarf estimated during primary eclipse is $L_{X}=3\times 10^{28}$~erg~s$^{-1}$, which is consistent with that of rapidly rotating ``saturated'' K and M dwarfs. 
\end{abstract}

\keywords{binaries: close --- X-rays: binaries --- stars: cataclysmic variables --- 
accretion, accretion disks}

\section{Introduction}

Short-period binaries composed of a white dwarf and late-type star are
of fundamental importance to astrophysics as they are progenitors of
cataclysmic variables (CVs) and novae, some of which likely evolve to
form type 1a supernovae.  Stellar evolution theory predicts that these
systems are the outcome of a common envelope evolutionary phase in
which frictions leads to a rapid spiraling down of the orbit of an
initial wider binary \citep{Paczynski:76}.   
At the conclusion of the common envelope phase, the orbit of the system is expected to decay further as a result of angular momentum lost to gravitational radiation and to the magnetized wind of the unevolved component \citep{Kraft.etal:62,Paczynski:67,Verbunt.Zwaan:81}.  As their separation decreases,
the Roche lobe of the secondary is reduced in size and is eventually 
completely filled, initiating the onset of mass transfer onto the white dwarf.

The timescale for initiation of mass transfer in a post-common envelope binary depends critically on the angular momentum loss rate.   While angular momentum loss through gravitational radiation is theoretically well-understood for the stars in CVs, there is no comprehensive theory of spin-down through magnetized winds that is expected to dominate the evolution of systems with periods above the CV period gap of $\ga 3$hr.    It depends on the wind mass loss rate and the large scale stellar magnetic field \citep[e.g.][]{Weber.Davis:67,Mestel:68,Kawaler:88}.  Both are extremely difficult to measure for late-type stars and are especially uncertain at the very rapid rotation rates of close binaries where magnetic proxies such as X-ray emission show saturation and supersaturation effects that appear to affect spin-down rates \citep[e.g.][]{Ivanova.Taam:03, Barnes:03b}.

In this paper, we present {\it XMM-Newton} observations of the
intriguing DA+M2-4 3.6hr eclipsing binary  QS Vir (formerly more commonly known as EC~13471-1258) 
that was first discovered as a white dwarf--M dwarf binary in the Edinburgh-Cape blue object survey \citep{Kilkenny.etal:97}.  QS~Vir has gained considerable recent attention as a valuable diagnostic of close binary evolution and angular momentum loss.  Based on an extensive
analysis of high-speed and multi-colour photometry, together with UV
HST STIS and visible light spectroscopy, \citet[][see also \citealt{Kawka.etal:02}]{O'Donoghue.etal:03}
concluded that the system filled its Roche lobe and was probably
undergoing very weak mass transfer and suggested the system might be a
hibernating cataclysmic variable.  It was detected in the ROSAT All Sky Survey (RASS) at $0.14\pm 0.02$~count~s$^{-1}$ \citep{Voges.etal:99} implying an inconspicuous X-ray luminosity $L_X=5\times 10^{29}$~erg~s$^{-1}$---much too low for an accreting  CV.  More recently, \citet{Qian.etal:10} inferred the presence of a giant planet with mass $6.4 M_{Jupiter}$ in a 7.86~yr orbit based on long term period variations,  although \citet{Parsons.etal:10} ruled this out from more complete eclipse monitoring.   Instead they find a third body in a highly elliptical orbit best fits the more complete data, noting that this explanation for the secular orbit change is also not without problems.    \citet{Ribeiro.etal:10} found evidence for material possibly associated with wind accretion within the Roche lobe of the white dwarf, while \citet{Parsons.etal:11} interpreted phase-specific absorption signatures as a stellar prominence originating from the M~dwarf.  The latter 
also found  the white dwarf to be slowly rotating, indicating a pre-cataclysmic evolutionary status.

Here, we report the discovery of deep, narrow X-ray eclipses 
locating the dominant source of X-rays on the white dwarf rather than
the corona of the late-type star and implying ongoing 
accretion.  
QS~Vir has one of the lowest accretion
rates of known non-magnetic systems.  This, together with its
eclipsing nature and proximity (48~pc; \citealt{O'Donoghue.etal:03}), renders
QS~Vir an interesting object for the study of CV evolution.

\section{Observations and Data Analysis}
\label{s:obs}

QS~Vir is an eclipsing binary with an inclination of $74\pm 2
^{\circ}$ and an orbital period of $P_{orb} = 0.15074 \pm 0.00004$ days.  The component 
spectral types are classified as DA
and M2--4 with masses $M_{wd} = 0.78 \pm 0.04\;
M_\odot$, $R_{wd} = 0.011 \pm 0.01\; R_\odot$ and $M_{md} = 0.43 \pm
0.04\; M_\odot$, $R_{md} = 0.42 \pm 0.02\; R_\odot$, respectively
\citep{Kawka.etal:02,O'Donoghue.etal:03,Howell.etal:10,Ribeiro.etal:10,Qian.etal:10}.

QS~Vir was observed using the instruments on board the
\emph{XMM-Newton} satellite \citep{Jansen.etal:01}.  Both RGS detectors
\citep{den_Herder.etal:01} 
operated in ``Spectroscopy'' mode, while the
EPIC-pn 
\citep{Struder.etal:01} 
and both MOS 
\citep{Turner.etal:01} 
cameras operated in ``Small window'' mode using a medium filter.  The OM
telescope 
\citep{Mason.etal:01} 
was used in ``Fast Mode'' with the $U$
filter, but a hardware glitch placed the source outside of the
photometry window and severely compromised the data.  A summary of the
observations is reported in Table~\ref{tab:obs}.  Here we concentrate on the EPIC observations; RGS spectra were of low signal-to-noise ratio, although a number of weak emission lines attributed to O~VIII, Fe~XVII, Ne IX an Ne X could be discerned.

\subsection{X-ray and Optical Photometry}
\begin{figure}
\epsscale{1.00}
\plotone{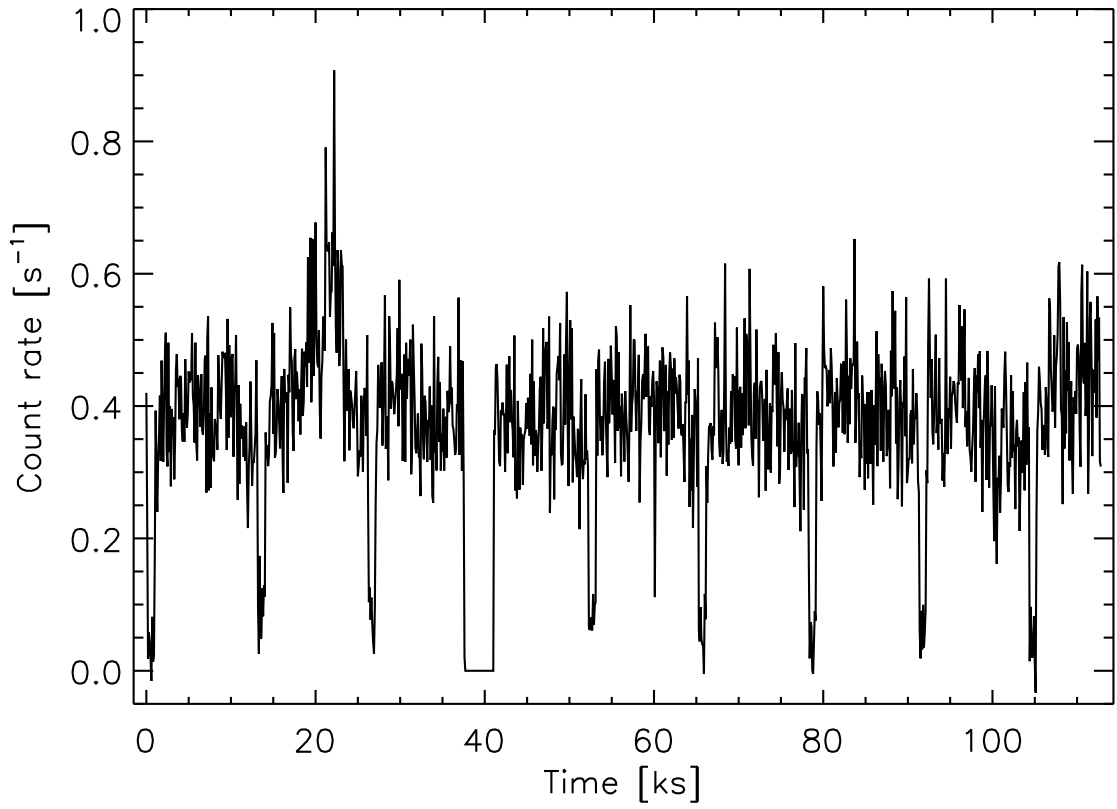}
\caption{EPIC-pn background-subtracted light curve illustrated with a 100 sec bin size as a function of time in seconds since the beginning of the observation.\label{fig:lc_pn}}
\end{figure}

Data were reduced with the \emph{XMM-Newton} Science Analysis System
software version 7.0.0, updated with the latest calibration files. The
EPIC-pn background-subtracted light curve with a 100s bin size is showed
in Fig.~\ref{fig:lc_pn}.  Apart from the sizable flare 20~ks into the
observation, which is likely due to coronal activity on the red dwarf,
the most conspicuous feature of the light curve is the presence of
deep eclipses of duration $\sim 800$s and a period of about 13~ks.  
This eclipse period and duration matches
that found by \citet{O'Donoghue.etal:03} in their UV-optical study of
QS~Vir, and we attribute the eclipses to the UV-bright white dwarf passing
behind the red dwarf.  At mid-eclipse, the X-ray emission declines to 
about 15\%\ of its uneclipsed intensity, indicating that $\sim
85$\%\ of the X-ray signal originates from the vicinity of the white
dwarf and dominates the quiescent coronal X-ray intensity of the
M~dwarf.

OM frames were analyzed using custom software designed to extract the residual spill-over signal 
in the photometry window from point spread function wings.  While not photometric, sufficient signal was recovered to verify that the X-ray and $U$-band eclipses coincided with each other.

\begin{figure}
\epsscale{1.00}
\plotone{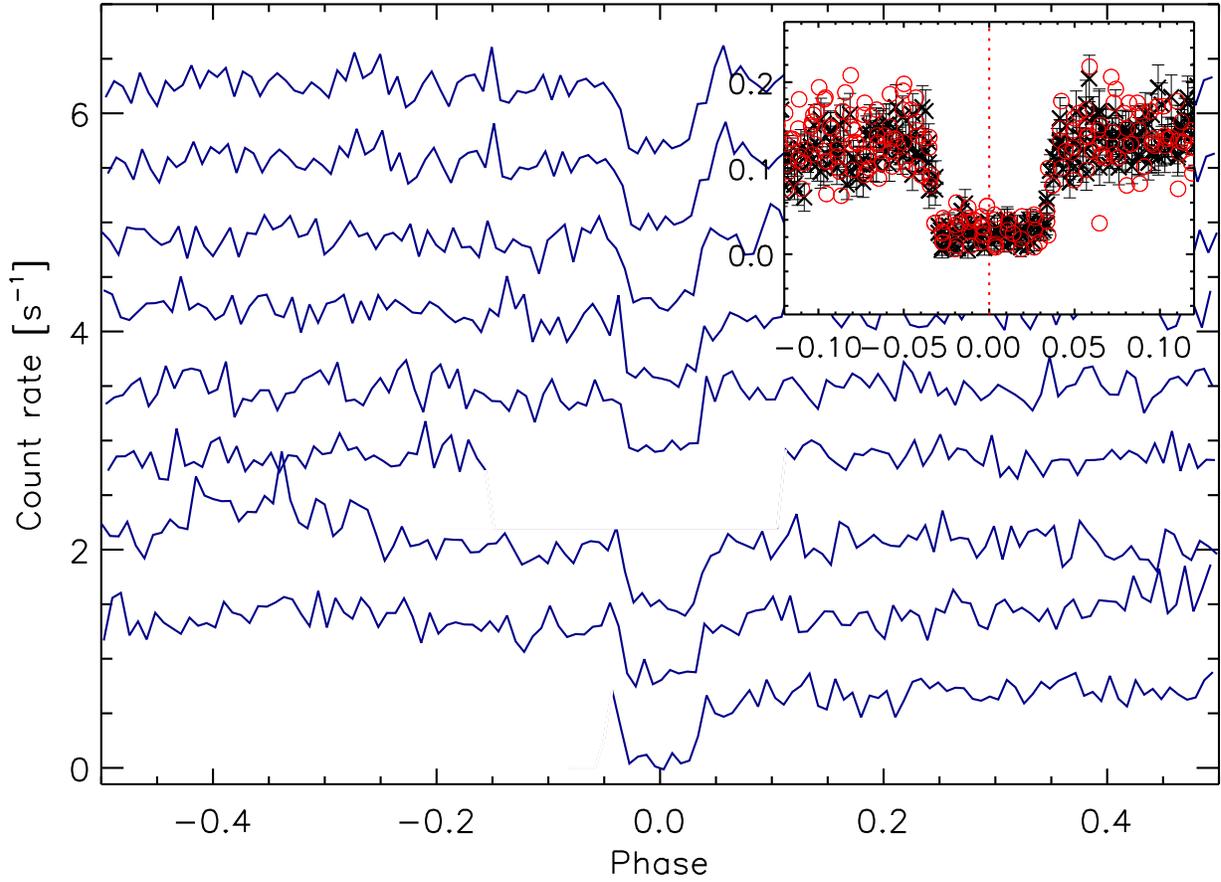}
\caption{EPIC-pn background subtracted lightcurve phase-folded according to the optical ephemeris of \citet{Qian.etal:10}.  Each phase is offset upward with respect to the previous one by 0.5~count~s$^{-1}$. \label{fig:lc_pn_cycles}}
\end{figure}

The EPIC-pn light curve phase-folded on the ephemeris for optical 
mid-eclipse derived from observations between 1992-2002 by \citet{O'Donoghue.etal:03} succeeded in co-aligning eclipses in phase space.  However, the X-ray eclipse preceded the optically-derived one by about 120s.  This is a much larger offset than expected based on the period drift apparent in Figure~3c of \citet{O'Donoghue.etal:03}, 
indicating that the orbital period had shortened somewhat in the  interval between the ephemeris epoch and our 2006 observations.   This secular period shortening was confirmed recently by the later optical observations of \citet{Qian.etal:10} and \citet{Parsons.etal:10}.   The ephemerides of these works are similar for the epoch of our observations and adopting arbitrarily that of the former we find our X-ray and $U$-band eclipses are centered on the expected primary eclipse.  The phase-folded X-ray light curve based on this ephemeris is illustrated in 
Fig.~\ref{fig:lc_pn_cycles}.  

Each cycle of the phase-folded X-ray light curve was fitted with a trapezoidal profile to locate the times of ingress and egress.  No significant variability in the eclipse mid-time or  duration are present, and the time of second to third contact is $\tau_{23} = 795\pm 40$s, in reasonable agreement with the value of $835\pm 1$s obtained from optical photometry by \citet{O'Donoghue.etal:03}.  This indicates that the eclipsed X-ray source is close to or coincident with the white dwarf photosphere. 

\subsection{EPIC Spectroscopy}

\begin{figure}
\epsscale{0.45}
\includegraphics[totalheight=0.85\textheight,angle=270]{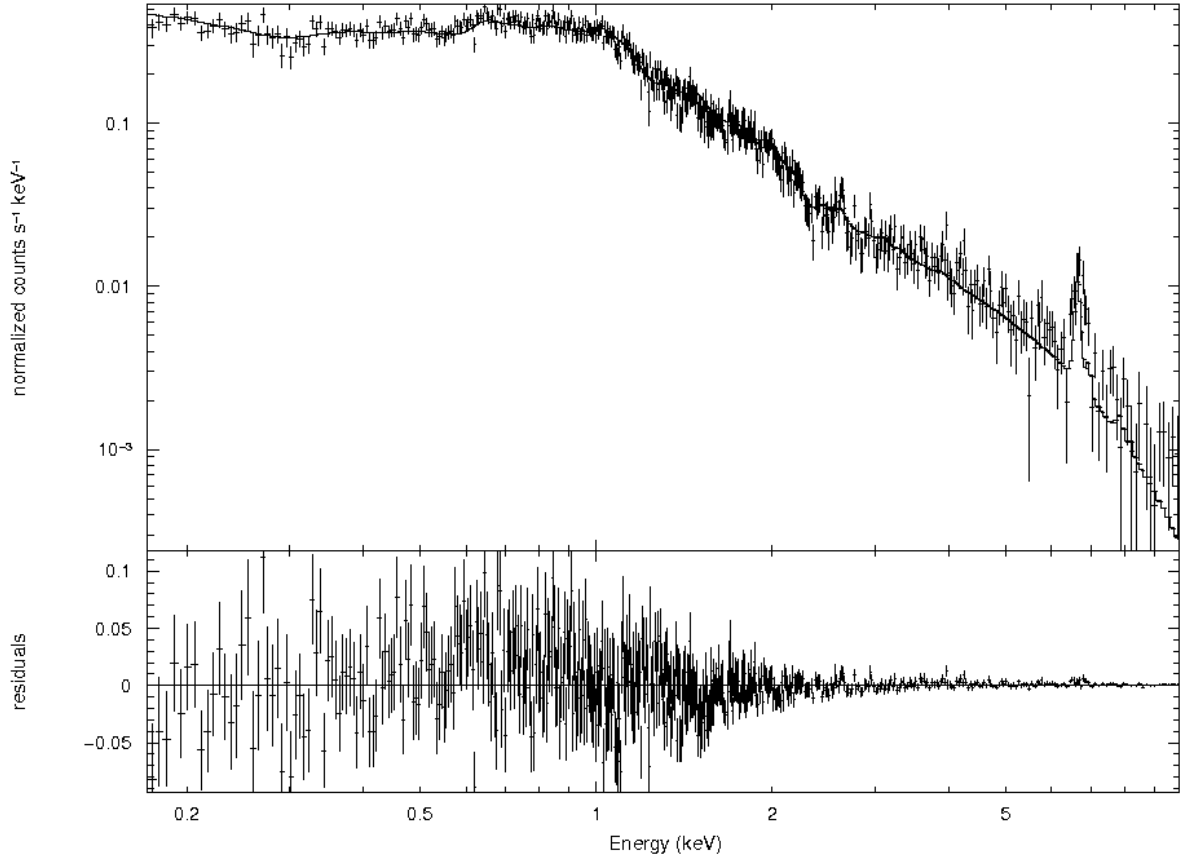}
\caption{EPIC-pn spectrum together with the best-fit MKCFLOW cooling flow model.\label{fig:pn_fit_flow}}
\end{figure}

EPIC pn source photons were extracted from a circular region with a $35\arcsec$ radius, while background was estimated from a surrounding annulus with an area of approximately ten times the source region.  Intervals showing particularly high background were excluded from the analysis.  The extracted EPIC pn spectrum, rebinned such that each channel contains at least 35 counts, is illustrated in Figure \ref{fig:pn_fit_flow}.

\cite{Mukai.etal:03} and \cite{Pandel.etal:05} have found that the X-ray spectra of CVs fall into two broad categories: optically-thin spectra resembling cooling flows; and spectra with a hard continuum exhibiting signatures of a photoionized plasma.  \cite{Mukai.etal:03} noted that the difference is likely to be associated with the accretion rate per unit area, with photoionization spectra belonging to magnetic systems with channeled accretion streams covering a small surface area, and optically-thin spectra coming from the accretion boundary layer of non-magnetic systems.   

Cursory inspection of the EPIC spectrum of QS~Vir reveals an optically-thin form not unlike that seen from active stellar coronae.   This is reinforced by the few emission lines that could be discerned from the RGS spectra.   We performed parameterized spectral modeling within the {\sc xspec} program version 12 \citep{Dorman.Arnaud:01}, using multi-temperature optically-thin plasma radiative loss models and a cooling flow model.   A fixed absorption of $N_{H}= 10^{18}$~cm$^{-2}$, corresponding to negligible interstellar X-ray absorption for this nearby object, was adopted for both models.  For the multi-temperature plasma, we employed a two-temperature MEKAL model (a m\'elange of the models of \citealt{Mewe.etal:85} and \citealt{Liedahl.etal:95}) in which the temperatures were allowed to vary freely but the metallicity parameter was forced to assume the same value for both components.  The best-fit model had a minimum value of the reduced $\chi^2$ statistic of  $\chi^{2}_{0} = 1.42$, with 520 degrees of freedom (dof).
For the cooling flow, we employed the isobaric flow model with emission measure verses temperature scaled inversely to the cooling time at that temperature \citep{Mushotzky.Szymkowiak:88} implemented as the MEKAL-based MKCFLOW model.  No redshift was included, and $\chi^{2}_{0} = 1.23$ with 520 dof was obtained with metallicity, minimum and maximum temperatures, and normalization as free parameters.  The parameter estimation results are reported in Table~\ref{tab:fit} and the best-fit model is illustrated in Figure~\ref{fig:pn_fit_flow}.  At face value, the cooling flow model describes the data slightly better than the two-temperature one, though we caution that systematic uncertainties in the model and instrument response are not included in the analysis.

\section{Discussion}

\subsection{Accretion}

The detection of clear, sharp X-ray eclipses phased with the optical eclipses provides an unambiguous indication of ongoing accretion in the QS~Vir system.   From the EPIC spectrum, we find the total X-ray luminosity in the 0.5--10~keV range to be $L_{X} = 2.1\times10^{29}$~erg~s$^{-1}$.  The source exhibits non-flaring variability at a level of about 20\%.  Based on the parameters of \citet{O'Donoghue.etal:03}, the bolometric luminosity of the M~dwarf is $L_{bol}=5.6\times 10^{31}$~erg~s$^{-1}$, and the ratio of X-ray to bolometric luminosity is $L_X/L_{bol}=5.4\times 10^{-4}$---comparable with the canonical empirical ``saturation'' value of $10^{-3}$ found for the most active stars \citep[e.g.][]{Pizzolato.etal:03} and typical of the pre-cataclysmic binaries and rapidly rotating M dwarfs in the solar neighborhood studied by \citet{Briggs.etal:07}.

If QS~Vir is a semi-detached system, the accretion rate obtained from the cooling flow model, $\dot{M}=(1.69\pm 0.05) \times 10^{-13}M_\odot$~yr$^{-1}$, is among the lowest ever found for a CV.   \citet{Lynden-Bell.Pringle:74} showed that  approximately half of the gravitational energy of the accreting gas is liberated through optical and UV radiation in a viscously-heated accretion disk and 
the other half is dissipated largely in the form of X-rays in a ``boundary layer'', where the disk material is decelerated from its Keplerian velocity to the rotation velocity of the white dwarf.  The expected X-ray luminosity from a disk boundary layer is then  $L_{BL}\simeq GM_{wd}\dot{M}/2R_{wd}$.  Using $R_{wd} = 0.011 R_\odot$ and $M_{wd} = 0.78 M_\odot$ \citep{O'Donoghue.etal:03},  the boundary layer luminosity is $ L_{BL} = 7\times10^{29}$~erg~s$^{-1}$, which is reasonably consistent with that observed.   If the radiative dissipation of viscous heating amounts to a similar luminosity in the UV and optical, it would be inconspicuous in comparison to the white dwarf luminosity of $L_{wd}= 1.7\times 10^{31}$~erg~s$^{-1}$
($T_{eff}=14220$~K; \citealt{O'Donoghue.etal:03}) as expected.
 

\subsection{Roche lobe overflow or just a wind?}

\subsubsection{Difficulties with Roche lobe overflow}

Accreting gas could originate either from the wind of the red dwarf, or from Roche Lobe overflow.  \citet{O'Donoghue.etal:03}, \citet{Ribeiro.etal:10}, and \citet{Parsons.etal:11} all found evidence for material within the Roche lobe of the white dwarf.  \citet{O'Donoghue.etal:03} suggest this might be associated with an accretion stream, while \citet{Ribeiro.etal:10} 
favored wind accretion because photometric modeling suggested the M dwarf did not fill its Roche Lobe.  \citet{Parsons.etal:11} suggested the presence of significant prominence material.  

An additional problem with Roche lobe overflow for QS~Vir is that this should lead to formation of an accretion disk.  At low accretion rates a disk can be more optically-thin and still render relatively strong disk emission lines, 
but no observable signatures of a disk have been detected. 
Since our observed mass accretion rate exceeds the total solar wind mass loss rate ($\sim 2\times 10^{-14} M_\odot$) by an order of magnitude, one crucial question to decide between accretion mechanisms is whether the mass loss rate of the M dwarf component of QS~Vir is sufficient to supply this flow through a wind alone.

\subsubsection{Insufficient wind supply?}

QS~Vir is in some respects similar to the ``pre-polars''---magnetic ($B > 10^6$G) pre-CVs that
do not appear to fill their Roche lobes yet show signs of accretion activity at low mass transfer rates.  Accretion on these systems is thought to be through magnetic capture of the secondary wind with a wind accretion efficiency of the order of 100\%\ (e.g., \citealt{Schwope.etal:02,Schmidt.etal:05}; see also, e.g., \citealt{Li.etal:94,Webbink.Wickramasinghe:05}).  Accretion rates in the range $\dot{M}=5\times10^{-14}$--$3\times 10^{-13}\,M_\odot\,$yr$^{-1}$ with no apparent dependence on orbital period or secondary spectral type have been found for several systems  
\citep[e.g.][]{Schwope.etal:02,Schmidt.etal:05,Schmidt.etal:07,Vogel.etal:11}.   These rates are significantly higher than wind accretion on non-magnetic pre-CV systems where magnetic capture is not expected to operate.  Total mass loss rates for M dwarfs in close non-contact binaries with non-magnetic white dwarfs of $\sim 10^{-14}$--$10^{-15} M_\odot$~yr$^{-1}$ have been estimated based on the gravitational settling time of wind-accreted metals \citep[e.g.][]{Debes:06,Tappert.etal:11}.
\citet{Schmidt.etal:07} argue that the pre-polar accretion rates could provide the 
first realistic measurements of stellar mass loss at the cool end of the main sequence.  Single M dwarf mass loss rates have otherwise been notoriously difficult to pin down; see, e.g., \citet{Wargelin.Drake:02} for a detailed discussion.     \cite{Wargelin.Drake:02} used astrospheric change exchange X-ray emission to place an upper limit of $\dot{M} \leq 3\times 10^{-13}M_\odot$~yr$^{-1}$ for the moderately active M5.5 dwarf Proxima, while \citet{Wood.etal:02} found $\dot{M} \leq 4\times 10^{-15}M_\odot$~yr$^{-1}$ from Ly$\alpha$ observations.   Both of these are reasonably consistent with the rates estimated for pre-polars.

The wind mass loss rate required to feed the observed $2\times 10^{-13}M_\odot$~yr$^{-1}$  accretion rate on QS~Vir depends on its accretion efficiency.   
In absence of detailed magnetohydrodynamic accretion simulations this efficiency must be considered unknown.  Exploratory MHD wind models for detached pre-CV systems by \citet{Cohen.etal:12} find accretion efficiencies depend quite sensitively on the WD and donor magnetic field strengths and on their orbital separation.
We can rule  out QS~Vir being a pre-polar with a strongly magnetized WD based on the 
absence of any Zeeman broadening or splitting of the Balmer lines or cyclotron emission features in the spectra illustrated by \citet{O'Donoghue.etal:03}, \citet{Ribeiro.etal:10} and \citet{Parsons.etal:11}.   The cores of the Balmer lines in QS Vir are dominated by emission, which would obscure any small Zeeman splittings and defines an upper limit to the possible field strength.  Based on the $\pm 5$~\AA\ width of the emission cores in the spectra illustrated in Figure~1 of \citet{Parsons.etal:11}, and using Equation~11 from \citet{Wickramasinghe.Ferrario:00} for a white dwarf temperature of 14200~K \citep{O'Donoghue.etal:03}, we estimate $B < 700$kG.  Adding some allowance for Stark broadening hiding a slightly higher field, we find a conservative limit of $B< 1$MG.  Such a limit is also consistent with the absence of cyclotron emission features, which are prominent in pre-polars with even lower accretion rates than QS~Vir \cite[e.g.][]{Schmidt.etal:07}.   \citet{Cohen.etal:12} did find one pre-CV configuration for a WD field $B=10^5$G in which the accretion rate approached 100\%\ of that of a single M star wind by an efficient ``siphoning'' of plasma from the hemisphere facing the WD, but this finding needs to be confirmed by more detailed simulations.  
The field would need to be of order $10^7$G to capture all the wind of the M dwarf, and the  accretion efficiency is then unlikely to be close to 100\%\ for QS~Vir.
  
An order of magnitude estimate of the pure hydrodynamic accretion efficiency can be made from the \citet{Bondi.Hoyle:44} recipe
\begin{equation}
\dot{M}_{acc}=\frac{4\pi G^2{M_{wd}}^2\rho(r)}{v^3}
\label{e:bondi}
\end{equation}
where the wind density, $\rho(r)$, as a function of radial distance $r$ in the vicinity of the white dwarf at orbital separation, $a$, for a wind mass loss rate $\dot{M}$ at velocity $v$ is 
\begin{equation}
\rho= \frac{\dot{M}}{4\pi v {R_{rd}}^2}
\label{e:density}
\end{equation}
Combining Eqns.~\ref{e:bondi} and \ref{e:density}, and assuming the wind velocity is approximately the escape velocity, $v\approx \sqrt{2 G M_{rd}/R_{rd}}$, (as is the case for the solar wind), the accretion rate is 
\begin{equation}
\dot{M}_{acc}=\left(\frac{R_{rd}M_{wd} }{2 a M_{rd}}\right)^2 \dot{M}
\end{equation}
where the first term on the right is the accretion efficiency. The orbital separation, $a=[P_{orb}^2 G(M_{wd}+M_{rd})/4\pi^2]^{1/3}$, from the parameters in \S\ref{s:obs} is $1.26R_\odot$ and the accretion efficiency is 0.09.  
Considering the accretion rate derived for QS~Vir from the one-dimensional cooling flow model is likely a lower limit,  the wind mass loss rate required to feed the accretion is then at least $\dot{M}\sim 2\times 10^{-12}M_\odot$~yr$^{-1}$---an order of magnitude larger than implied for the M dwarfs of the pre-polars.  For a typical velocity of 600~km~s$^{-1}$, this wind would require $\sim 1$\%\ of the stellar bolometric luminosity to drive which we consider uncomfortably high.

\citet{Qian.etal:10} interpreted a residual period decrease of QS Vir with respect to their ephemeris solution that included a giant planet as angular momentum loss from a wind with an even higher mass loss rate of $7.2\times 10^{-11}M_\odot$~yr$^{-1}$.  
Massive winds, up to $10^{-10}M_\odot$~yr$^{-1}$, were argued earlier by \citet{Mullan.etal:92} based on thermal escape arguments together with weak evidence of infrared wind free-free emission (see also \citealt{Badalyan.Livshits:92}).   More recently, \citet{Vidotto.etal:11} predicted a similarly massive wind for the M4 dwarf V374~Peg from magnetohydrodynamic modeling, with a ram pressure 5 orders of magnitude larger than the solar wind.   However, such massive winds can be readily dismissed on energetic grounds.  For a terminal velocity of $\sim 500$~km~s$^{-1}$, the kinetic power of the wind suggested by \citet{Qian.etal:10} is $6\times 10^{30}$~erg~s$^{-1}$---an implausibly large fraction (10\%) of the M dwarf luminosity.   Similarly, the wind of \citet{Vidotto.etal:11} corresponds to an unphysical $\sim 10^{32}$~erg~s$^{-1}$ of kinetic energy when compared with 
the underlying stellar bolometric luminosity  of $L_{bol}=3.6\times 10^{31}$~erg~s$^{-1}$ \citep{Morin.etal:08}.   \citet{Lim.White:96} had pointed out  earlier that such massive winds would be opaque to radio emission which is commonly observed from active late-type stars, and deduced that mass loss rates can be no larger than 1-2 orders of magnitude above the solar rate. 
For QS~Vir, the need for such a massive wind is obviated by the additional period measurements of \citet{Parsons.etal:10}, who instead invoked a third body with a minimum mass of $0.05M_\odot$ in an elliptical orbit, together with the activity-related quadrupole coupling mechanism of \citet{Applegate:92}, to explain the period variations.  \citet{Parsons.etal:10} noted evolutionary difficulties for the presence of such a body, though it and the secular period evolution are not of any obvious relevance for understanding the pre-CV evolutionary status, parameters and accretion rate of QS~Vir.  A further unexplored connection cannot of course be ruled out.

The kinetic power of the solar wind, $\sim 10^{27}$~erg~s$^{-1}$, is similar to the mean observed radiative power of the corona \citep[e.g.][]{Peres.etal:00}.  \citet{Schwadron.etal:06} argued that  the energy available to drive the solar wind scales with the basal open magnetic field strength.  Combining this with an observed, almost linear, correlation between X-ray radiance and surface magnetic flux \citep{Pevtsov.etal:03}, the expectation would be for stellar wind mass loss rate to scale with X-ray luminosity.   \citet{Wood.etal:02} estimated mass loss rates for a handful of different stars of different activity using astrospheric Ly$\alpha$ absorption, and found $\dot{M}\propto F_X^{1.15\pm {0.20}}$, where $F_X$ is the surface X-ray flux.  Scaling to the observed surface X-ray flux of QS~Vir yields $\dot{M}\sim 5\times 10^{-12}M_\odot$~yr$^{-1}$---consistent with our estimated accretion requirement.  However, there also exist no measurements for stars at the saturated activity level of QS~Vir and \citet{Wood.etal:02} caution against  extrapolating the relation.  Proxima also does not support the relation, with a wind upper limit from Ly$\alpha$ absorption lying an order of magnitude below the predicted value.

In summary, for an assumed wind accretion efficiency of the order of 10\%, the preponderance of more direct assessments of the likely wind mass loss rates of active M dwarfs like the secondary of QS~Vir, in particular those for the pre-polars, suggest it would not be sufficient to feed the observed accretion rate.  

\subsubsection{Chromospheric mass supply in a rare evolutionary phase?}

An alternative to catastrophically energetic wind or accretion efficiency requirements is surface activity that does not lead to mass loss in single stars.  \citet{Parsons.etal:11} suggested prominence activity could be responsible for material within the Roche lobe of the white dwarf.  Such activity is stochastic and might be expected to result in highly variable X-ray emission.   The X-ray luminosity found here from XMM-Newton observations, $(2.1\pm 0.4)\times 10^{29}$~erg~s$^{-1}$, is about a factor of two lower than that measured in the RASS ($(4.9\pm 0.7)\times 10^{29}$~erg~s$^{-1}$)---perhaps suggestive of moderate accretion variability though flaring on the M dwarf cannot be ruled out. 

A more constant, lower lying, source is 
upward chromospheric spicule and mottle flow.  In the solar chromosphere and transition region this flow corresponds to a mass flux two orders of magnitude larger than is launched in the solar wind \citep[e.g.][]{Campos:84,Tsiropoula.Tziotziou:04}.   The flow has a typical scale height of $\sim 5000$~km and falls back to the solar surface, but might more easily succeed in escaping the Roche surface of a star on the verge of Roche lobe overflow to become readily unbound to the M dwarf and available to be swept up by the lobe of the white dwarf.  Such a mass flow source would only become available for a near lobe-filling system.  Following, e.g., \citet{King.Kolb:95}, the timescale for evolution through such a chromospherically-fed phase is 
\begin{equation}
\tau_{ch}=\frac{h_{ch}}{R_{rd}} \frac{J}{2 \dot{J}}
\label{e:timescale}
\end{equation}
where $h_{ch}=kT_{ch}{R_{rd}}^2/\mu GM_{rd}$ is the chromospheric scale height, $J$ the system angular momentum and $\dot{J}$ its time derivative.  For a chromospheric temperature of $10^4$~K, the ratio $h_{ch}/R_{rd}=4.3\times 10^{-4}$, though for spicules this height is an order of magnitude greater.  Adopting the \citet{Skumanich:72}-based \citet{Verbunt.Zwaan:81} spin-down prescription modified by \citet{Rappaport.etal:83}
\begin{equation}
\dot{J}=-3.8\times 10^{30} M_{rd} {R_\odot}^4 \left(\frac{R_{rd}}{R_\odot}\right)^\gamma \left(\frac{2\pi}{P_{orb}}\right)^3 
\end{equation}
with $\gamma=2$ and where 
\begin{equation}
J=\left(\frac{2\pi}{P_{orb}}\right) \frac{a^2 M_{rd}M_{wd}}{M_{rd}+M_{wd}}
\end{equation}
we find a timescale $\tau_{ch}=5\times 10^4$~yr for the chromospheric scale height and $5\times 10^5$~yr for spicules.  The implication of this short timescale is that chromospheric mass supply would occur only in a rare, fleeting episode of evolution.



\section{Conclusions and the nature of QS Vir}



X-ray spectrophotometry of QS~Vir betrays ongoing accretion at a rate of $2\times 10^{-13}M_\odot$~yr$^{-1}$.  The slow rotation of the white dwarf diagnosed by \citet{Parsons.etal:11} argues against a hibernating system, and QS~Vir is most likely at the first onset of significant accretion.  QS~Vir then represents a valuable object to probe the very earliest stages of CV evolution. 
At face value, the observed accretion rate implies Roche lobe overflow and a semi-detached system.
A pure wind accretion alternative would imply a wind mass loss rate of $\dot{M}\sim 2\times 10^{-12}M_\odot$~yr$^{-1}$ for a Bondi-Hoyle accretion efficiency of 10\%.  For a typical velocity of 600~km~s$^{-1}$, this wind would require 1\%\ of the stellar bolometric luminosity to drive.  While plausible, we consider such a massive and energetic wind unlikely.  If QS~Vir is still detached, as the analysis of  \citet{Ribeiro.etal:10} implies, accreted mass could be supplied by prominence activity \citep{Parsons.etal:11}, though such stochastic activity would likely lead to highly variable X-ray emission.  The X-ray luminosity observed by XMM-Newton is a factor of two lower than that seen in the RASS and weakly supports this idea.  We further speculate here that upward chromospheric mass flow analogous to solar spicules and mottles that escapes the Roche surface to be swept up by the white dwarf would present a more constant mass supply.  Such a supply would become available for about million years or so prior to Roche lobe overflow and QS~Vir would then be in a very interesting, fleeting evolutionary phase.  


\acknowledgments

We are indebted Boris G\"ansicke for very useful comments and discussion of low accretion rate CV-like systems.
MM was supported by NASA Grant NNG06GD34G during the course of
this work.  JJD and VK were funded by NASA contract NAS8-39073 to the {\it
Chandra X-ray Center} (CXC) and
thank the CXC director, Harvey Tananbaum, and the CXC science team
for advice and support.


\clearpage

\begin{deluxetable}{ccccc}
\tablewidth{0pt}
\tablecaption{Summary of the observations\label{tab:obs}}
\tablehead{
\colhead{Instrument} & 
\colhead{Exp. Time} & 
\colhead{Mode} &
\colhead{Filter} &
\colhead{Obs. Start Time}\\
\colhead{} & 
\colhead{[ksec.]} & 
\colhead{} &
\colhead{} &
\colhead{}
}
\startdata
RGS1 & 109.5 & Spectroscopy & \nodata & 2006-01-25@18:13:37 \\
RGS2 & 109.5 & Spectroscopy & \nodata & 2006-01-25@18:13:42 \\
MOS1 & 109.0 & Small Window & medium & 2006-01-25@18:14:51 \\
MOS2 & 108.9 & Small Window & medium & 2006-01-25@18:14:51 \\
pn & 109.3 & Small Window & medium & 2006-01-25@18:20:24 \\

OM & 17.6 & Fast Mode & U & 2006-01-25@18:20:22 \\
\multicolumn{1}{c}{''}& 17.6 & Fast Mode & U & 2006-01-26@00:53:50 \\
\multicolumn{1}{c}{''}& 21.8 & Fast Mode & U & 2006-01-26@07:57:18 \\
\multicolumn{1}{c}{''}& 21.7 & Fast Mode & U & 2006-01-26@14:28:41 \\
\multicolumn{1}{c}{''}& 21.7 & Fast Mode & U & 2006-01-26@20:57:59 \\

\enddata
\end{deluxetable}

\clearpage

\begin{deluxetable}{lc|lc}
\tablewidth{0pt}
\tablecaption{Summary of the parameter estimation results\label{tab:fit}}
\tablehead{
\multicolumn{2}{c}{COOLING FLOW} & 
\multicolumn{2}{c}{2-TEMPERATURE} \\ 
\colhead{Parameter} &
\colhead{Best-fit} &
\colhead{Parameter} &
\colhead{Best-fit}
}
\startdata
N$_{H}$ (cm$^{-2}$)\tablenotemark{a} & 10$^{18}$ &  N$_{H}$ (cm$^{-2}$)\tablenotemark{a}&  10$^{18}$ \\
low~T (keV) & $0.08\pm0.15$ &  kT$_1$ (keV) & $0.67\pm0.01$ \\
high~T (keV) & $7.24\pm0.17$ &  kT$_2$ (keV)  & $3.58\pm0.08$  \\
Metallicity\tablenotemark{b} & $0.44\pm0.05$ &  Metallicity\tablenotemark{b} & $0.48\pm0.03$ \\
$\dot{M}$~($10^{-13} M_\odot$/yr)  & $1.694\pm0.046$ &  EM$_1$ ($10^{52}$~cm$^{-3}$)\tablenotemark{c}  & $0.26\pm 0.03$ \\
 & &  EM$_2$  ($10^{52}$~cm$^{-3}$)\tablenotemark{c} & $2.08\pm 0.04$ \\

\enddata
\tablenotetext{a}{Parameter held fixed.}
\tablenotetext{b}{Expressed as a fraction of the solar metal abundances listed by \citet{Anders.Grevesse:89}}
\tablenotetext{c}{Assuming a distance of 48~pc \citep{O'Donoghue.etal:03}}
\end{deluxetable}

\clearpage

\clearpage

\clearpage





\end{document}